\documentclass[twoside]{article}
\usepackage{fleqn,espcrc2}

\newcommand{\be}{\begin{equation}}
\newcommand{\ee}{\end{equation}}
\newcommand{\bea}{\begin{eqnarray}}
\newcommand{\eea}{\end{eqnarray}}

\begin{document}

\author{E.P.~Likhtman \address{All-Russian Institute of Scientific and 
Technical Information, \\ Usievich Str. 20, Moscow 125315, Russia
\\
e-mail: likh@polly.phys.msu.su}}
\title{Around SuSy 1970}

\maketitle

Let me turn back to look at the past, and tell you about the history
of  construction of the first supersymmetric model. The papers I will
shortly review have been for the most part rather inaccessible. A part of
the  results obtained is presented only in my Ph.D. thesis. I will
also mention  some lost opportunities and delusions.

I graduated from the Physical Department of Moscow State University
obtaining the diploma with honors in 1968. At that time, I was a co-author of
two publications on some plausible charge-parity breaking effects
in atomic nuclei. I dreamed of solving more ambitious problems, but
this required  becoming a postgraduate student. However my diploma
supervisor, Yuri Shirokov, could not  arrange my postgraduate studies
--- neither at the Physics Department of Moscow State University, where he
was a Professor, nor at the Mathematics Institute of the Academy of
Science, where he was a staff member. Therefore, he recommended me to
Yuri Golfand who was working in the Theory Department of the Lebedev
Physical Institute of the Academy of Sciences.

In the late 1960's --- early 70's, the staff members of the
Theoretical Department were famous Soviet physicists such as Igor Tamm,
Vitaliy Ginzburg,and  Andrey Sakharov. At that time Golfand was still
preparing his doctoral thesis \cite{b1}. A moderate height, a rapid gait
and a charming well-wishing smile were to level down the differences between
our ages and positions. He showed me several commutation and
anticommutation relations between the operators of  momentum, angular
momentum and some spinors and explained to me that their consistency was
verified by Jacobi identities. These commutation relations provided
the basis for the science, which  in six years was called the SuperSymmetry.

At the first stage I had to establish whether the proposed algebra was
unique or  whether there were some alternatives. In order to do this, it was
necessary to solve a system of equations for the algebra structure
constants, which follow from the Jacobi identities. I restricted
myself to the set of four complex spinor charges and found that there
were four versions of such algebras: two of them are now known as 
${\cal N}=1$ and ${\cal N}=2$ superalgebras, in two others the momentum
did not commute with spinor charges (\`a la de-Sitter algebra). These
results were obtained in 1968 but were published later only in 1972,
because their relation to physics was not immediately clear. We chose
the simplest (${\cal N}=1$) supersymmetric algebra for further
investigations.

Twice a week, before seminars of the Theoretical Department, I showed
the result of my calculations to Golfand, and we discussed them. At
that time I used  text-books on different aspects of applying 
group theory to physics, but I did not read any new publications.
Later, when I was preparing my Ph.D. thesis, I had to write a review
of modern literature on the subject, and I faced difficulties again.
Just recently I have read the historical review ``Revealing the path to the
superworld" by Marinov in the Golfand  Memorial Volume \cite{b2} 
with great interest. 
I don't know whether Golfand knew about Felix Berezin's papers or any
other publications on the subject. Maybe, he believed, that he had
already informed me about all that was necessary for our work. By all
his visual appearance Golfand gave me, and probably not only me, an
idea that, contrary to others, we worked on a very serious and
important problem. At the same time, he seemed to have good relations
with all staff members of the Theoretical Department. He liked jokes
and banters very much, in particular about  gaps in my education.

The main problem to be solved was to relate the constructed algebra to
quantum field theory. Nobody knew whether such a relation 
existed, and if yes, whether its representations were finite
dimensional. I use here the word ``algebra" instead of ``group" for the
following reason. We introduced the notion of supercoordinates and
considered the group with Grassmann parameters and established the
supercoordinate transformations under spinor translations. But we
didn't guess to expand the superfield with respect to the Grassmann
variables and to establish the relation between the superfield and a
set of the usual fields forming the supermultiplet. The notion of the
Superfield was not presented in the published papers \cite{b3,b4}, 
because for us it 
did not provide a mechanism for constructing a superinteraction; it was
presented in my Ph.D. thesis only. The way to success, as it seems
to be now, was not so elegant and, therefore, more laborious. I began
to seek the representations of the algebra in terms of creation and
annihilation operators.

As long as we had to do with the algebra, the main problem was to
properly handle the gamma-matrices and not to forget to change the
signs in certain places of the Jacobi identities while dealing
with anticommutators. When we turned to field theory, we found it
very difficult to get accustomed to the fact that both bosons and
fermions are in the same multiplet and even more difficult to imagine 
how  they transform into each other under spinor translation. 
There was no way to get an answer from a crib as I used to do at
my exams on the Marx-Lenin philosophy. Golfand encouraged me: ``A cat
may look at a king." The only thing that was clear from the very
beginning was that all particles in the multiplet have equal masses,
but this fact didn't give optimism.

At last, in 1969 the superspin operator and two irreducible
representations of algebra (chiral multiplet of spins zero and one
half and vector multiplet of spins zero and one half and one) were
constructed and following general properties of irreducible
representations were established:
\begin{itemize}
\item   First, the  maximum spin in any irreducible representation differs
   from minimum one by not more than one. This conclusion was obtained
   by expanding the group operator (not the superfield) in Taylor
   series with respect to Grassmann parameters of spinor translations.
   The expansion turned out to have a cut-off and therefore the
   irreducible representations were finite!

\item   The second discovered property was that the number of bosonic and
   fermionic degrees of freedom in every multiplet was the same, and,
   as a consequence, the total vacuum energy of bosons and fermions was
   equal to zero.
\end{itemize}
I cite here the Lebedev Physical Institute preprint \#41, 1971 (see Appendix), 
of which the 
only reader was probably Misha Shifman. I wanted very much to publish
the obtained results immediately, in 1969, but Golfand
believed that they would attract no attention. Moreover, he didn't
want to lose time on the manuscript preparation. Therefore, I began
constructing an interaction of the found multiplet.

At that time the Academy of Science was receiving letters from
inventors with requests to make examination of their projects. Post
graduate students had to make sense of them and to give a conclusion.
I was given a project of a rocket, where liquid moved inside the
rocket in a closed tube: straight in one direction and in the zigzag
fashion in the opposite direction. The author of the project believed
that due to the relativistic effects a force has to arise and push the
rocket forward. I could not point out to the author that this
contradicted the momentum conservation law, because the author,
evidently, did not know about it. I had to spend a lot of time in order to
find mistakes in his considerations. After this case I was afraid
to find myself in a similar situation, in the position
of luckless inventor.

The psychological barrier associated with the fermi-bose mixture was
broken, but some technical difficulties arose. The method of
constructing interactions was to write down a general form of
spinor generators of the algebra not only in terms of the second but
also the third powers of the fields. Consequently, the number of unknown
constants in front of different combinations of fields was about a dozen.
Today I solve about eight hundred of nonlinear equations with
ease and have problems with the computer only if the number of equations
exceeds a thousand. To solve the equations in 1970 I used pen and the
reverse side of  blueprints of old drawings. It is easy to solve a
problem from a text-book, when you know, that the answer does exist and
the only thing to do is to find it. When one of equations contradicted
the others I didn't know whether this was the result of arithmetic
mistakes or whether the problem had no solution at all. My postgraduate
term went to finish and I had to think about my Ph.D. thesis and
future job. Finally, when constants obtained from one of the equations
appeared to satisfy the others, I took it as a miracle. Moreover, the
unknown constants at the fourth powers of fields in the spin
translation operators could be set to zero. The system was solved, the
first superinteraction was constructed. Now it is known as  massive
supersymmetric electrodynamics.

The time came to draw the results up. I was going to write down one
big consistent paper, but Golfand took another decision. Time of 
publication in our journals was very large, and he decided to write
down a short paper for JETP Letters \cite{b3}. Golfand cut down my manuscript
without any pity to fit the required volume. Deleted fragments were
scattered around in other publications. At the same time I constructed
the selfinteraction for the vector multiplet. I succeeded in constructing
 only
the trilinear part of the interaction, therefore the result was presented
only in my Ph.D. thesis. The title of the thesis was so complicated,
that during the defense of my Ph.D. thesis in September 1971, the scientific
secretary of our institute faced  difficulties in reading it aloud.
The title seemed to reflect our understanding of the problem. We
erroneously believed, that constructing  supersymmetric interaction
through the power series expansion of group generators in the coupling
constant was related to  specific features of supersymmetry.
Nevertheless, summarizing the results of three years' work I conclude that
under the continual support by Golfand, I had solved the problem
formulated by him, that is I demonstrated the quantum field
realization of supersymmetry by a specific example.

What stimulated Golfand to formulate the problem? It is clear that it
was not any specific experimental result such as constant speed of
light or the approximate equality of the neutron and proton masses.
There were also no questions of overcoming some internal
contradictions in field theory as it took place in constructing
general relativity or the Weinberg-Salam model. I think that he was
inspired by  numerous positive results which were obtained by applying
one or another symmetry in physics during the 20th century.

At the end of 1971 I got a job at the Physics Department of All-Union
Institute of Scientific and Technical Information and hardly found
time to continue my work on fermi-bose symmetry.  Golfand's
position was even worse: he lost his job and became unemployed.

I really hated  divergences in  quantum field theory, and the
hypothesis came to me that the cancelation of the vacuum energies
divergences for free fields, that I found, would take place for
interacting fields as well. Knowing coupling constants it was easy to
establish that the one-loop mass divergences of bosonic fields in
constructed model will not be quadratic, but logarithmic as for
fermionic field. Golfand was not excited by this result because he
believed it happened just by chance. Maybe his intuition left him, but
maybe his mind was occupied by completely different non-scientific
problems. Considering the action of spinor transforms on $S$-matrix, I came
to the conclusion, that the cancelation of the divergences was not
accidental and had to take place in higher orders. However, the
divergences did not disappear completely. What if the
nonrenormalizable theory became renormalizable one? Unfortunately,
with my methods for constructing models there were no chances to treat
nonrenormalizable models.

In 1974 Igor Tyutin told me about the Wess-Zumino paper which I have
read with great interest. The linear realization of representations
owing to the presence of
 auxiliary fields, covariant derivatives with respect on
Grassmann variables --- it was very beautiful, and allowed one to
construct the superinteraction rather easily. Integration over
Grassmann variables, used in the paper by Salam-Strathdee, finally set 
both types of arguments of superfields on equal footing. The
postulates of this integration were formulated by Felix Berezin in
1965. Using these techniques I constructed two models with the Abelian
and non-Abelian massive vector fields and proved their
renormalizability. At the same time I also noted that the proof of
renormalizability could be carried out in another way, using the
standard mechanism of spontaneous gauge symmetry breaking. The
question of any relation of supersymmetry to high energy 
physics remained open.

After the  publication of these two papers \cite{b5},
I practically stopped my
work on supersymmetry because of strong competition and 
the absence of any
support. I began to seek my own unique topic. A small number of
assumptions and the simplicity of constructions, together with
nontrivial and maybe not very realistic (at first glance) results ---
that was the main impression, which I gained from my collaboration
with Golfand. On the other hand, my ability to operate the personal
computers allows me to try to solve problems, while I do not even
know about the existence of their solutions. I heard that to seek a black
cat in a dark room is very risky business, especially in its absence.
Nevertheless, today I am trying to study numerically the model of
Born-Infeld electromagnetic field interacting with a membrane of
finite size and with a string boundary, where it is beforehand known
that the divergences of electric and magnetic energies cancel each
other. But this cancelation is not the final aim. It appears that
some combination of mass, electric charge and magnetic momentum is
independent on two unknown dimensional constants of the model on the
one hand and, on the other hand, is related to well-known observable
value --- the fine structure constant. At the present time, the
solutions  obtained 
slowly converge with the growth of the number of the lattice
cells. Maybe this points  to the absence of solutions, or maybe it is
necessary to arrange the lattice better, but maybe it is necessary to
seek the unknown symmetry of the equations of motion to use it for
preliminary analytical analysis.

In conclusion I would like to thank the organizers of the conference
{\em Thirty Years of Supersymmetry}
for giving me the opportunity to share my vision of thirty-year old
events, and for financial support. I would like to express my 
gratitude to Misha Shifman for comments to my notes in the 
Golfand Memorial Volume \cite{b2}. These comments
made some aspects of the Soviet life in the 1970's more palpable for
the Western readers. And, the last but not the least, Yuri Golfand 
remains for me
not only the man who taught me a trade, but the man who made me love the
risky business --- not to be afraid to go deep in the forest by the paths
nobody walked on before. Thank you for your attention.

\twocolumn[
\section*{APPENDIX}

At the suggestion of Misha Shifman I present below a {\em verbatim}
translation of my paper written in 1970 which circulated as FIAN preprint 
No.~41 (1971). The final authorization for issuing this preprint was obtained 
on April 12, 1971.

\vspace{3mm}

\begin{center}
LEBEDEV PHYSICS INSTITUTE OF THE 
USSR ACADEMY OF SCIENCES\\
\vspace{0.5cm}
Theory Department
\end{center}

\begin{flushright}
Preprint No. 41\\
Moscow 1971
\end{flushright}

\vspace{0.1cm}

\begin{center}
{\bf\large IRREDUCIBLE REPRESENTATIONS OF THE EXTENSION OF THE ALGEBRA 
OF GENERATORS OF THE POINCAR\'E GROUP BY BISPINOR GENERATORS}\\
\vspace{0.4cm}
{\large E.P.~Likhtman}
\end{center}
\vspace{0.5cm}
]

\section{INTRODUCTION}

In Ref.~1 a special extension of the algebra ${\cal P}$ of the generators 
of the Poincar\'e group was considered. The extension was performed by virtue 
of introduction of the generators of the spinorial translations 
$W_\alpha$ and $\bar W_\beta$, 
\renewcommand{\theequation}{1\alph{equation}}
\begin{eqnarray}
\lefteqn{ [M_{\mu\nu}, M_{\sigma\lambda} ]_- = 
   i (\delta_{\mu\sigma} M_{\nu\lambda} + \delta_{\nu\lambda} 
   M_{\mu\sigma}} \nonumber\\
&&\mbox{}- \delta_{\mu\lambda} M_{\nu\sigma}  - \delta_{\nu\sigma} M_{\mu\lambda} ); 
   \nonumber\\
\lefteqn{ [P_\mu, P_\nu]_-=0;} \nonumber\\
\lefteqn{ [M_{\mu\nu}, P_\lambda ]_- =
 i(\delta_{\mu\lambda} P_\nu -\delta_{\nu\lambda} P_\mu);} \nonumber\\
\lefteqn{ [M_{\mu\nu}, W ]_- = \frac{i}{4} [\gamma_\mu,\gamma_\nu] W; } \\ 
\lefteqn{ \bar W = W^+ \gamma_0 ; } \nonumber\\ 
\lefteqn{ [ W, \bar W]_+=\gamma_\mu^+ P_\mu; } \nonumber\\ 
\lefteqn{ [W,W]_+ = 0; \qquad [P_\mu, W]_- = 0 ; }\\ 
\lefteqn{\gamma^\pm_\mu=s^\pm \gamma_\mu; \qquad s^\pm = \frac{1}{2} 
(1\pm \gamma_5), \qquad \gamma_5^2 = 1.} \nonumber
\end{eqnarray}

\renewcommand{\theequation}{\arabic{equation}}
\setcounter{equation}{1}

The spinorial indices are omitted, while the expressions of the type 
$d_\mu d_\nu$ are to be understood as $d_0 d_0 - d_1 d_1 - d_2 d_2 - d_3 d_3$ 
hereafter. In the same work a realization of this algebra was constructed in
which a Hamiltonian describes interactions of quantum fields. This example shows
that algebra (1) imposes rigid constraints on the form of the quantum field
interaction. In constructing this example we used two linear irreducible
representations of the algebra (1). Their definition was not given in Ref.~1. In
this work I present the definition of these representations of algebra (1), and
build other representations. The properties of these representations are
investigated.

\section{THE SPACE OF STATES AND INVARIANT SUBSPACES}

In order to determine in which space the representations of algebra (1) act note
that the algebra ${\cal P}$ is its subalgebra. Therefore, any representation of
algebra (1) is also a representation of the algebra ${\cal P}$. The spaces in
which these representations act coincide. However, irreducible representations
of algebra (1) are reducible with respect to ${\cal P}$. This reducible
representation splits in several irreducible representations of ${\cal P}$ and
one and the same irreducible representation can be involved several times. To
distinguish them we will introduce the number $\chi$ of the irreducible
representations.

Of physical interest are those representations of (1) which can be reduced to
(several) representations of ${\cal P}$ characterized by mass and spin.
therefore, the basis vector of the space in which we build representations of
(1) can be written as follows:
\be 
|\kappa, p_\lambda, j, m, \chi \rangle ,
\ee
where $\kappa$ is the mass, $p_\lambda$ is the spatial momentum, $j$ stands for
the spin, $m$ is its projection on the $z$ axis, and, finally, $\chi$ is the
number of the irreducible representation of ${\cal P}$. 

In the space with the basis vectors (2) there are subspaces invariant under the
action of the operators from the algebra in Eq.~(1). According to Schur's
lemma \cite{cit2}, in order to find invariant subspaces it is necessary to find
invariant operators which, by definition, commute with all operations of algebra
(1). It is easy to observe that the operator $P_\mu^2$ does have this property.
Therefore, the space of which the basis vectors correspond to particles of one
and the same mass $\kappa$ will be the invariant subspace. The spins of the
vectors of this invariant subspace cannot be all the same, since the square of
the spin operator is not an invariant operator of algebra (1),
\begin{displaymath}
[ \Gamma_\mu^2 , W ]_-\neq 0; \qquad (\Gamma_\mu = \frac{1}{2} 
\varepsilon_{\mu\nu\lambda\sigma} M_{\nu\lambda} P_\sigma) . 
\end{displaymath}
Instead, the invariant operator of algebra (1) is $D_\mu^2$ where
\begin{eqnarray*}
\lefteqn{ D_\mu = \Gamma_\mu + \frac{1}{2} (\bar W \gamma_\mu W 
- \frac{P_\mu P_\nu}{P_\sigma^2} \bar W \gamma_\nu W),} \\
\lefteqn{ P_\sigma^2 = \kappa^2 > 0 .} 
\end{eqnarray*}
Other invariant operators in algebra (1) are likely to be absent.

{\em A priori} it is not exactly known which vectors (2) form the basis
of the irreducible representation of (1), since the properties of the operator
$D_\mu^2$ are not known. Besides the fact that the mass of all states is the
same, one can assert that the difference between the maximal and minimal spins
is the irreducible representation of (1) does not exceed 1. In other words,
irreducible representations of (1) contain no more than 3 distinct spins.
Otherwise, by consecutively  applying the operators from algebra (1) to the
state vectors with spin $j_1$ one could obtain a vector with a nonvanishing
projection on the state vectors with spin $j_2$, $|j_2-j_1|>1$. To see that this
is impossible note that the consecutive action of the operators from algebra (1)
in the most general case can be represented as a polynomial in these operators.
Due to the (anti)commutation relations (1) the form of this polynomial is
\bea
\lefteqn{\sum C_{\mu\nu,\ldots;\lambda\ldots}^{\alpha,\ldots;\beta,\ldots}} 
\nonumber\\
\!\!\!& \times &\!\!\! 
M_{\mu\nu} \! \times \! \ldots\! \times \! P_\lambda \! \times \! \ldots 
\! \times \! W_\alpha \! \times\! \ldots \! \times \! \bar W_\beta 
\! \times\! \ldots , \nonumber
\eea
where $C_{\mu\nu,\ldots;\lambda\ldots}^{\alpha,\ldots;\beta,\ldots}$ are
numerical coefficients.

The product of any number of operators $M_{\mu\nu}$ and $P_\lambda$ does not
change the spin of the state. The number of operators $W$ (and $\bar W$) in each
term can be equal to one or two since if it is larger than 2 then the product
vanishes because of the anticommutation relation in Eq.~(1). By the same reason
the product $W_\alpha W_\beta\neq 0$ only provided that $\alpha\neq\beta$. Such
operator does not change the spin of the state as well. Thus, an invariant
subspace can contain only the spins $j$, $j+\frac{1}{2}$, $j+1$. 

In order to build the irreducible representation of algebra (1) it is necessary
to find the matrix elements of the operators from algebra (1) between these
states.

\section{EQUATIONS FOR THE REDUCED MATRIX STATES}

The matrix element of the operator of conventional translations 
can obviously be
written as 
\bea 
\lefteqn{\langle \kappa, p_\lambda, j, m, \chi | P_\mu | 
   \kappa, p'_\lambda, j', m',\chi'\rangle} \nonumber\\
&=& (\delta_{\mu 0}\sqrt{\kappa^2+p_\lambda^2} +\delta_{\mu\lambda} p_\lambda) 
\nonumber\\
&\times&   
\delta(p_\lambda - p'_\lambda) \delta_{jj'} \delta_{mm'} \delta_{\chi\chi'} ,
\eea
where $\lambda=1,2,3$. The operator $M_{\mu\nu}$ will have its 
conventional form
as well. We will be interested in the matrix elements of the operators
$W_\alpha$ and $\bar W_\beta$, which, due to Eq.~(1), are diagonal with respect
to $\kappa$ and $p_\lambda$. One can readily convince oneself that the 
operators
$s^- W$ and $\bar W s^+$ satisfy the trivial commutation relations. Therefore,
we will limit ourselves to investigating such representations in 
which~\footnote{Relaxing this requirement would require, of necessity,
the introduction of an indefinite metric \cite{cit3}. } 
\begin{displaymath} s^- W = \bar s^+ =0. 
\end{displaymath}
Without loss of generality let us choose the representation of the $\gamma$
matrices with a diagonal $\gamma_5$. Then the operators $W_\alpha$ and 
$\bar W_\beta$ become two-component. Moreover, knowing the transformation law of
spinors under the Lorentz transformations, and assuming that $\kappa > 0$, let
us pass to the reference frame with $p_\lambda=p'_\lambda = 0$. In this reference
frame one could apply the Wigner-Eckart theorem \cite{cit4} according to which 
\bea 
\lefteqn{\langle \kappa, 0, j, m, \chi | s^+ W_\alpha | 
   \kappa, 0, j', m',\chi'\rangle } \nonumber\\
&=& \left( \begin{array}{ccc} 
        j & \frac{1}{2} & j' \\
        m & \alpha & -m' 
  \end{array} \right) 
  (-1)^{j'-m'} \nonumber\\
&\times&\sqrt{(2j+1)(2j'+1)} \langle j\chi | f | j'\chi'\rangle ; 
  \nonumber\\
\lefteqn{\langle \kappa, 0, j, m, \chi | \bar W s^-_\beta | 
   \kappa, 0, j', m',\chi'\rangle } \nonumber\\
&=& \left( \begin{array}{ccc} 
        j' & \frac{1}{2} & j \\
        m' & \beta & -m 
  \end{array} \right) 
  (-1)^{j-m} \nonumber\\
&\times&\sqrt{(2j+1)(2j'+1)} \langle j\chi | f^+ | j'\chi'\rangle , 
\eea
where $\left( \begin{array}{ccc} j & \frac{1}{2} & j' \\ m & \alpha & -m' 
\end{array} \right)$ are the Wigner symbols, $|j-j'|=\frac{1}{2}$, 
$\langle j\chi | f | j'\chi'\rangle$ are the reduced matrix elements, 
$\sqrt{(2j+1)(2j'+1)}$ is a convenient normalization factor. The representation
(4) ensures the correct commutation relation with the momentum operator and the
operator of the spatial rotations. In order to satisfy other commutation
relations of algebra (1), we substitute (3) and (4) in (1b) and exploit the
formulae of summation of the $3j$ symbols in the spin projections \cite{cit4}, 
\begin{eqnarray*}
\lefteqn{\sum\limits_{m'} (-1)^{j'+m} } \\
&\times& \left( \begin{array}{ccc}
        j_1 & j_2 & j' \\ m_1 & m_2 & m' \end{array} \right) 
        \left( \begin{array}{ccc}
        j_3 & j_4 & j' \\ m_3 & m_4 & -m' \end{array} \right) \\
&=& \sum\limits_{JM} (-1)^{2j_4+J+M} (2J+1) \\
&\times& \left\{ \begin{array}{ccc}
        j_1 & j_2 & j' \\ j_3 & j_4 & J \end{array} \right\} 
        \left( \begin{array}{ccc}
        j_3 & j_2 & J \\ m_3 & m_2 & M \end{array} \right) \\
&\times& \left( \begin{array}{ccc}
        j_1 & j_4 & J \\ m_1 & m_4 & -M \end{array} \right) ,
\end{eqnarray*}
where $\left\{ \begin{array}{ccc} j_1 & j_2 & j' \\ j_3 & j_4 & J \end{array} 
\right\}$ is the $6j$ symbol. As a result of this substitution we obtain, after
performing the summation, 
\bea 
\lefteqn{\sum\limits_{JM j' \chi'} (-1)^{2j''+J+M} (2J+1)} \nonumber\\
&\times& \left\{ \begin{array}{ccc}
        j & j' & J \\ \frac{1}{2} & \frac{1}{2} & j' \end{array} \right\} 
        \left( \begin{array}{ccc} 
        \frac{1}{2} & \frac{1}{2} & J \\ -\beta & \alpha & M \end{array} \right)
\nonumber\\
&\times& \left( \begin{array}{ccc}
        j & j'' & J \\ m & -m'' & -M \end{array} \right) \nonumber\\
&\times& (-1)^{m+\alpha-j'} \sqrt{(2j+1)(2j''+1)(2j'+1)^2} \nonumber\\
&\times& \Big( \langle j\chi|f|j'\chi'\rangle 
   \langle j'\chi'|f^+| j''\chi''\rangle \nonumber\\
&+& (-1)^{J-j+j''} \langle j\chi|f^+| j'\chi'\rangle 
   \langle j'\chi'|f| j''\chi''\rangle \Big) \nonumber\\
&=& \kappa \delta_{j j''} \delta_{m m''} \delta_{\chi \chi''} .
\eea 
(An analogous formula is obtained upon substitution of Eq.~(4) in the
anticommutation relation $[W,W]_+=0$.) Next we use the values of the $6j$
symbols \cite{cit4}, 
\begin{eqnarray*} 
\lefteqn{\left\{ \!\! \begin{array}{ccc}
        j_1 & j_2 & j_3 \\ \frac{1}{2} & j_3-\frac{1}{2} & j_2+\frac{1}{2} 
        \end{array} \!\! \right\}  = (-1)^{j_1+j_2+j_3} }\nonumber\\
&\times& \left[ 
\frac{(j_1+j_3-j_2)(j_1+j_2-j_3+1)}{(2j_2+1)(2j_2+2)(2j_3)(2j_3+1)} 
\right]^\frac{1}{2} ; \\
\lefteqn{\left\{ \!\! \begin{array}{ccc}
        j_1 & j_2 & j_3 \\ \frac{1}{2} & j_3-\frac{1}{2} & j_2-\frac{1}{2} 
        \end{array} \!\! \right\} = (-1)^{j_1+j_2+j_3}} \nonumber\\
&\times&\left[ 
\frac{(j_1+j_2+j_3+1)(j_2+j_3-j_1)}{(2j_2)(2j_2+1)(2j_3)(2j_3+1)} 
\right]^\frac{1}{2} .
\end{eqnarray*} 
\renewcommand{\theequation}{6\alph{equation}}
\setcounter{equation}{0}
Then Eq.~(5) takes the form 
\bea 
\lefteqn{\sum\limits_{j' \chi'} \left(\frac{2j'+1}{2}\right) 
   \Big( \langle j\chi|f|j'\chi'\rangle \langle j'\chi'|f^+| j''\chi''\rangle}
   \nonumber\\
&+& \langle j\chi|f^+| j'\chi'\rangle \langle j'\chi'|f| j''\chi''\rangle \Big)
\nonumber \\   &=& \kappa \delta_{j j''} \delta_{\chi \chi''} ; \\
\lefteqn{\sum\limits_{j' \chi'}(-1)^{j'} 
   \Big( \langle j\chi|f|j'\chi'\rangle \langle j'\chi'|f^+| j''\chi''\rangle}
   \nonumber\\
&-& \langle j\chi|f^+| j'\chi'\rangle \langle j'\chi'|f| j''\chi''\rangle \Big) 
\nonumber\\   
&=& 0, \quad \mbox{for } j\neq 0; \\
\lefteqn{\sum\limits_{j' \chi'} 
   \langle j\chi|f|j'\chi'\rangle \langle j'\chi'|f^+| j''\chi''\rangle } 
   \nonumber\\
&=& 0, \quad \mbox{except for } j=j''=0.
\eea
These are the equations that were our goal. Their solution will allow us to find
the explicit form of $s^+ W$ and $\bar W s^-$. Note that  Eq.~(6) describes the
representation in which the invariant operator $D_\mu^2$ need not necessarily be
proportional to the unit operator, i.e. we get reducible representations of
algebra (1), generally speaking. 
\renewcommand{\theequation}{\arabic{equation}}
\setcounter{equation}{6}

\section{THE NUMBER OF PARTICLES IN THE REPRESENTATION OF ALGEBRA (1) AND SOME 
SOLUTIONS OF EQ.~(6) }

First of all, starting from Eq.~(6) I will deduce constraints on the number of
particles in the representation of algebra (1). To this end let us multiply
Eq.~(6a) by $(-1)^{2j} (2j+1)/2$ and sum over $j=j''$ and $\chi=\chi''$. After
this operation the left-hand side will vanish. To see that this is the case it is
sufficient to transpose two factors in the second term (which is justified since
this is inside the trace) and to use the fact that $(-1)^{2j}=(-1)^{2j'+1}$ 
(see Eq.~(4)). Then the second term will differ from the first one by the sign
only. The right-hand side of Eq.~(6a) must also vanish, and we obtain the
constraint on the number $n$ of the particles with spin $j$ in the
representation of algebra (1), 
\be 
\sum\limits_j (-1)^{2j} (2j+1) n_j = 0.
\ee

As is known \cite{cit5}, in relativistic quantum field theory, the particle
energy operator, being transformed to the normal form, contains an infinite term
which is interpreted as the vacuum energy. It is also known that the sign of
this term is different for particles subject to the Bose and Fermi statistics.
According to Eq.~(7), the representations of algebra (1) include particles with
different statistics, so that the number of the boson states is always equal to
that of the fermion states. From this it follows that the infinite positive
energy of the boson states is canceled by the infinite negative energy of the
fermion states in any representation of algebra (1).

After these preliminary remarks we proceed directly to solving Eq.~(6). Let us
try to find representation in which only particles with two distinct spin values
enter. In this case $j'$ in Eq.~(6) takes only one value, and the summation over
$j'$ is, in fact, absent. Using this fact, let us multiply Eq.~(6b) by 
$(-1)^{j'} (2j'+1)/2$ and add up the result with Eq.~(6a). Then on the
right-hand side one will find a nonsingular matrix acting in the space of the
variable $\chi$ (at fixed $j$, $j'$, and $j''$). On the left-hand side the
matrix will be nonsingular only if $j\neq 0$. (Cf. Eq.~(6c)). Therefore, the
representation of algebra (1) with two spins can contain only spin-0 and
spin-1/2 states. It is easy to check that in this case the simplest solution of
the system with $n_0=2$ ($\chi=1,2$) and $n_{1/2}=1$ ($\chi=1$) has the form 
\be 
\langle j\chi|f|j'\chi'\rangle = \sqrt{\kappa} \left( \begin{array}{cc|c}
        0 & 0 & 1 \\ 0 & 0 & 0 \\ \hline 0 & 1 & 0 \end{array} \right) ,
\ee
where the matrix on the right-hand side acts on the state 
\begin{displaymath}
\left(\begin{array}{c} a\\ b\\ \hline c\end{array}\right) .
\end{displaymath} 
The amplitudes $a$ and $b$ describe the spin-0 particles while $c$ describes the
spin-1/2 particle.

In the case of the three-spin representations, $j$, $j+1/2$, and $j+1$, the
lowest spin $j$ can be arbitrary. The simplest solution with $n_j=1$ ($\chi=1$),
$n_{j+1/2}=2$ ($\chi=1,2$), and $n_{j+1}=1$ ($\chi=1$) can be written in a form
analogous to (8), 
\be 
\langle j\chi|f|j'\chi'\rangle = \sqrt{\kappa} \left( \begin{array}{c|cc|c}
        0 & 0 & 1 & 0 \\ \hline 1 & 0 & 0 & 1 \\ 0 & 0 & 0 & 0 \\ 
        \hline 0 & 0 & -1 & 0 \end{array} \right) .
\ee

The representations (8) and (9) are irreducible. One can readily convince
oneself that this is the case even without calculating the eigenvalues of the
operator $D_\mu^2$. It is enough to observe that there exist no representations
with a lesser number of particles satisfying the necessary condition (7). The
question of existence/nonexistence of other irreducible representations of (1),
besides those found here, requires further investigation.

\section{SECOND-QUANTIZED RELATIVISTIC REPRESENTATIONS OF ALGEBRA (1) }

In this section we will show how one can pass from the representations (8), (9)
acting in the space with the basis vectors (2) to the relativistic-covariant 
form of these representations, acting in the space of the occupation numbers.
The operators of the algebra in such representations have to be expressible in
terms of the second-quantized free fields with equal masses but distinct spins.
Since relativistic equation for spin-1/2 particles describes both particles and
antiparticles, it is necessary  to introduce antiparticles in the representation
(7) . Then the operators of the algebra will be expressible in terms of
non-Hermitian free scalar fields $\varphi(x)$, $\omega(x)$, and a spinor field
$\psi_1(x)$. Let us show that the operator 
\bea 
W^o = s^+ W^o &=& \frac{1}{i} \int 
   \Big( \varphi^*(x) \stackrel{\leftrightarrow}{\partial_0} s^+ \psi_1(x)
\nonumber\\   
&+& \omega(x) \stackrel{\leftrightarrow}{\partial_0} s^+ \psi_1^c(x) \Big)
   d^3 x ,
\eea 
(where the superscript $o$ means that the operator is bilinear in the field
operators while the superscript $c$ means the charge conjugation) 
satisfies~\footnote{The operator $W$ is defined up to a phase factor, see
Ref.~3.} (anti)commutation relations (1). For instance, 
\bea 
\lefteqn{ [ W^o, \bar W^o ]_+ 
= i\int\!\!\!\!\int d^3 x \; d^3 y} \nonumber\\
&\times& \Big( \varphi^*(x) 
   \stackrel{\leftrightarrow}{\partial_{x_0}} \gamma_\mu^+ i \partial_{x_\mu} 
   D(x-y) \stackrel{\leftrightarrow}{\partial_{y_0}} \varphi(y) \Big) 
   \nonumber\\
&+& i\int\!\!\!\!\int d^3 x \; d^3 y \nonumber\\
&\times& \Big(\omega(x) 
   \stackrel{\leftrightarrow}{\partial_{x_0}} \gamma_\mu^+ i \partial_{x_\mu} 
   D(x-y) \stackrel{\leftrightarrow}{\partial_{y_0}} \omega^*(y) \Big) 
   \nonumber\\ 
&+& i\int\!\!\!\!\int d^3 x \; d^3 y \nonumber\\
&\times& \Big(\bar\psi_1(y) s^- 
   \times \stackrel{\leftrightarrow}{\partial_{y_0}} 
   D(y-x)\stackrel{\leftrightarrow}{\partial_{x_0}} s^+ \psi_1(x) \Big) 
   \nonumber\\ 
&+& i\int\!\!\!\!\int d^3 x \; d^3 y \nonumber\\
&\times& \Big(\bar\psi_1^c(y) s^- 
   \times \stackrel{\leftrightarrow}{\partial_{y_0}} 
   D(y-x)\stackrel{\leftrightarrow}{\partial_{x_0}} s^+ \psi_1^c(x)\Big)
   \nonumber\\ 
&=& -\int d^3 x \; \Big(\varphi^*(x) \partial_\mu  
   \stackrel{\leftrightarrow}{\partial_0} \varphi(x) \Big) \times \gamma_\mu^+ 
\nonumber\\   
&-&\int d^3 x \; \Big(\omega^*(x) \partial_\mu 
   \stackrel{\leftrightarrow}{\partial_0} \omega(x) \Big) \times \gamma_\mu^+
\nonumber\\
&+& \frac{i}{2} \int d^3 x \; \Big(\bar\psi_1(x) 
   \stackrel{\leftrightarrow}{\partial_0} \gamma_\mu^- \psi_1(x) \Big) 
   \times \gamma_\mu^+ \nonumber\\
&+& \frac{i}{2} \int d^3 x \; \Big(\bar\psi_1^c(x) 
   \stackrel{\leftrightarrow}{\partial_0} \gamma_\mu^- \psi_1^c(x) \Big) 
   \times \gamma_\mu^+ \nonumber\\
&=& \int d^3 x \; T_{\mu 0}(x) \times \gamma_\mu^+ .
\eea 

In this calculation I used the Fierz identity, the equations of motion, and the
commutation relations for free fields. Let us further note that the
energy-momentum tensor $T_{\alpha\beta}$ is nonsymmetric in the case of the
spinor field, generally  speaking. However, if one of the indices is zero, then 
\begin{displaymath}
T_{\mu 0} = T_{0 \mu},
\end{displaymath}
and the integral on the right-hand side of Eq.~(11) becomes the energy-momentum
tensor of the fields $\varphi(x)$, $\omega(x)$, and $\psi_1(x)$. All other
relations in (1) can be checked in a similar manner. The action of the operator
$W^o$ on the field operators is a linear transformation of these fields.
Schematically, one can write it as follows: 
\begin{eqnarray*} 
\lefteqn{\varphi(x) \to \psi(x) \to \omega(x) \to 0 ,} \\
\lefteqn{\omega^*(x) \to \psi^c(x) \to \varphi^*(x) \to 0 .}
\end{eqnarray*} 

Let us pass now to generalization of the representation (9) to cover the case of
the quantized fields. We will limit ourselves to the option that the lowest spin
is zero, while two spin 1/2 particles may be considered related by the operation
of the charge conjugation. Then the operators of algebra (1) in this
representation will be expressed in terms of a Hermitian scalar field 
$\chi(x)$, Hermitian vector transverse field $A_\mu(x)$, and a spinor field 
$\psi_2(x)$. This irreducible representation can differ from that in Eq.~(10) by
the mass of the particles and must differ by the eigenvalues of the invariant
operator $D_\mu^2$. The operator $W^o$ in this representation has the form 
\bea 
W^o = s^+ W^o &=& \frac{1}{i\sqrt{2}} \int \Big(\chi(x) 
   \stackrel{\leftrightarrow}{\partial_0} s^+ \psi_2(x) \nonumber\\
   &+& A_\mu 
   \stackrel{\leftrightarrow}{\partial_0} \gamma_\mu^+ \psi_2(x) \Big) d^3 x .
\eea 
One can verify Eq.~(12) in the same manner as Eq.~(10), 
\bea 
\lefteqn{ [s^+ W^o, \bar W^o s^-]_+ = 
-\frac{1}{2i} \int\!\!\!\!\int d^3 x \; d^3 y } \nonumber\\ 
&\times&   \Big(\bar \psi_2(y) s^- \times 
   \stackrel{\leftrightarrow}{\partial_{y_0}} D(y-x) 
   \stackrel{\leftrightarrow}{\partial_{x_0}} s^+ \psi_2(x) \Big) \nonumber\\
&+& \frac{1}{2i} \int\!\!\!\!\int d^3 x \; d^3 y \Big( \psi_2(y) 
   \gamma_\mu^+\times 
   \stackrel{\leftrightarrow}{\partial_{y_0}}  \nonumber\\
&\times& (\delta_{\mu\nu} +\frac{1}{\mu^2} 
   \partial_{y_\mu} \partial_{y_\nu} ) D(y-x) 
   \stackrel{\leftrightarrow}{\partial_{x_0}} \gamma_\nu^+ \psi_2(x) \Big)
   \nonumber\\
&-& \frac{1}{2} \int \Big(\chi(x) \stackrel{\leftrightarrow}{\partial_0} 
   \partial_\mu \chi(x) \Big) d^3 x \times \gamma_\mu^+ \nonumber\\
&-& \frac{1}{2} \int \Big(A_\mu(x) \stackrel{\leftrightarrow}{\partial_0}
   \partial_\alpha A_\nu(x) \Big) d^3 x \times 
   \gamma_\mu^+ \gamma_\alpha^- \gamma_\nu^+ \nonumber\\
&=& \gamma_\mu^+ P_\mu ,
\eea 
where $\mu\neq 0$ is the mass of the fields $\chi(x)$, $A_\mu(x)$, and
$\psi_2(x)$. In this representation the action of the operator $W^o$ on free
fields can be schematically depicted as 
\begin{displaymath} 
\psi_2^c(x) \ {\nearrow\atop\searrow} 
\begin{array}{c} \chi(x)\\  \\  A_\mu(x) \end{array}
{\searrow\atop\nearrow} \ \psi_2(x) \to 0 . 
\end{displaymath}
At intermediate stages the mass $\mu$  enters in the denominator in Eq.~(13).
Therefore, it cannot be set to zero. The limit $\mu\to 0$ can be realized by
abandoning the condition of transversality of the vector field and passing to the
diagonal pairing 
\begin{displaymath}
[A_\mu(x), A_\nu(y)]_-=-\frac{1}{2} \delta_{\mu\nu} D(x-y) . 
\end{displaymath}
In this case the field $\psi_2(x)$ becomes two-component ($s^+ \psi_2=0$). The
first term in intermediate calculations in~(13) becomes unnecessary and
disappears; therefore, there is no need in the introduction of the scalar field
$\chi(x)$. At $\mu=0$ the operator $W^o$ has the following form 
\bea 
W^o &=& s^+ W^o \nonumber\\
&=& \frac{1}{i\sqrt{2}} \int \Big( A_\mu(x) 
   \stackrel{\leftrightarrow}{\partial_0} \gamma_\mu^+ \psi_2(x) \Big) d^3 x .
\eea 

In Secs.~2--4 where I investigated the properties of the representations with
nonvanishing mass, it was shown that the numbers of the fermion and boson states
in the representations of algebra (1) coincide. Therefore, the operator 
$P_\mu^o$ is automatically representable in the normal form. This can be seen
also from the fact that the action of the operators $W^o$ and $\bar W^o$ on the
vacuum always yields zero, and 
\begin{displaymath} 
P_\mu = {\rm Tr} (\gamma_\mu^- [W^o, \bar W^o]_+ ) .
\end{displaymath}
Therefore, the representation (14) also possesses this property, while the
vector and spinor massless particles can only be in two states with the 
opposite chiralities.

\section{CONCLUSION}

Summarizing, we found several irreducible representations of algebra (1). In
these representations conventional fields are united in certain multiplets. A
question arises whether one can identify these multiplets with some observed
particles. In answering this question the main difficulty lies in the fact that
masses of all particles in the multiplet are identical, while spins are
different. Therefore, at present algebra (1) and its realizations must be
considered as just a model of a Hamiltonian formulation of quantum field theory.

In conclusion I express my deep gratitude to Yu.A.~Golfand for his constant
attention to my work and stimulating discussions.

\end{document}